\documentclass {article}
\usepackage{graphicx}

\begin{document}

\title{On the thermodynamic origin of the Hawking entropy and a measurement of the Hawking temperature}

\author{Michael Petri\thanks{email: mpetri@bfs.de} \\Bundesamt f\"{u}r Strahlenschutz (BfS), Salzgitter, Germany}

\date{{May 1, 2004}}

\maketitle

\begin{abstract}

In the spherically symmetric case the Einstein field equations
take on their simplest form for a matter-density $\rho = 1 / (8
\pi r^2)$, from which a radial metric coefficient $g_{rr} \propto
r$ follows. The boundary of an object with such an interior
matter-density is situated slightly outside of its gravitational
radius. Its surface-redshift scales with $z \propto \sqrt{r}$, so
that any such large object is practically indistinguishable from a
black hole, as seen from exterior space-time.

The interior matter has a well defined temperature, $T \propto 1 /
\sqrt{r}$. Under the assumption, that the interior matter can be
described as an ultra-relativistic gas, the object's total entropy
and its temperature at infinity can be calculated by microscopic
statistical thermodynamics. They are equal to the Hawking result
up to a possibly different constant factor.

The simplest solution of the field equations with $\rho = 1 / (8
\pi r^2)$ is the so called holographic solution, short "holostar".
It has an interior string equation of state. The strings are
densely packed, explaining why the solution does not collapse to a
singularity. The holographic solution has been shown to be a very
accurate model for the universe as we see it today in Ref[7].

The factor relating the holostar's temperature at infinity to the
Hawking temperature can be expressed in terms the holostar's
interior (local) radiation temperature and its (local)
matter-density, allowing an experimental verification of the
Hawking temperature law. Using the recent experimental data for
the CMBR-temperature and the total matter-density in the universe
measured by WMAP, the Hawking formula is verified to an accuracy
better than 1\%.

\end{abstract}

\section{\label{sec:Intro}Introduction:}

In \cite{petri/bh} several new exact solutions to the Einstein
field equations were derived. These solutions are characterized by
a spherical boundary membrane, consisting out of tangential
pressure.

One of the new solution turned out to be of particular interest.
The so called holographic solution is characterized by the
property that its boundary membrane carries a
stress-energy-content equal to its gravitating mass. The
membrane's pressure is equal to the pressure of the - fictitious -
membrane attributed to a black hole by the membrane paradigm. This
guarantees, that the holostar's action on the exterior space-time
is by all practical purposes identical to that of a black hole.

The holostar's geometric properties have been discussed
extensively in \cite{petri/hol}. The holographic solution has no
free parameters, yet it turned out to be an astoundingly accurate
description of the universe as we see it today.

The holostar's interior matter-state can be interpreted as a
collection of radially outlayed strings, attached to the
holostar's spherical boundary membrane \cite{petri/string}. The
interior strings are densely packed, their mutual transverse
separation is exactly one Planck area. This dense package
explains, why the holostar doesn't collapse to a singularity,
although its boundary membrane lies just roughly two Planck
coordinate distances outside of its gravitational radius.

Although the holostar's total interior matter-density has a
definite string character, at least part of the matter can be
interpreted in terms of particles. In this paper a simple
thermodynamic model for the interior matter state is explored,
which allows us to derive the Hawking entropy and temperature
relations for a spherically symmetric black hole by microscopic
statistical thermodynamics in the ideal gas approximation.

\section{\label{sec:holo:intro}A short introduction to the holographic solution}

The holographic solution is an exact solution to the Einstein
field equations with zero cosmological constant. The spherically
symmetric metric of the holographic solution has been derived in
\cite{petri/bh} :

\begin{equation}
ds^2 = g_{tt}(r) dt^2 - g_{rr}(r) dr^2 - r^2 {d\Omega}^2
\end{equation}

\begin{equation} \label{eq:A}
g_{tt}(r) = 1/g_{rr}(r) = \frac{r_0}{r}(1-\theta(r-r_h)) +
(1-\frac{r_+}{r})\theta(r-r_h)
\end{equation}

with

$$ r_h = r_+ + r_0$$
$$ r_+ = 2M$$

All quantities are expressed in geometric units $c=G=1$. For
clarity $\hbar$ will be shown explicitly. $\theta$ and $\delta$
are the Heavyside-step functional and the Dirac-delta functional
respectively. $r_h$ denotes the radial coordinate position of the
holostar's surface, which divides the space-time manifold into an
interior source region with a non-zero matter-distribution and an
exterior vacuum space-time. $r_+$ is the radial coordinate
position of the gravitational radius (Schwarzschild radius) of the
holostar. $r_+$ is directly proportional to the gravitating mass
$M = r_+/2$. $r_0$ is a fundamental length
parameter.\footnote{$r_0$ has been assumed to be roughly twice the
Planck-length in \cite{petri/bh, petri/hol}. The analysis in
\cite{petri/charge} indicates $r_0^2 \simeq 4 \sqrt{3/4}$ at low
energies. In this paper a more definite relationship in terms of
the total number of particle degrees of freedom at high
temperatures will be derived.}

The matter fields (mass density, principal pressures) of any
spherically symmetric gravitationally bound object can be derived
from the metric by simple differentiation (see for example
\cite{petri/bh}). For the discussion in this paper only the {\em
radial} metric coefficient $g_{rr}(r)$ is essential. In the
spherically symmetric case the total mass-energy density $\rho$
can be calculated {\em solely} from the radial metric coefficient.
For any spherically symmetric self gravitating object the
following general relation holds:

\begin{equation} \label{eq:r/grr}
{\left(\frac{r}{g_{rr}}\right)}' = 1 - 8 \pi r^2 \rho
\end{equation}

It is obvious from the above equation, that a matter-density $\rho
= 1 / (8 \pi r^2)$ is special. It renders the differential
equation for $g_{rr}$ homogeneous and leads to a strictly linear
dependence between $g_{rr}$ and the radial distance coordinate
$r$.

With $g_{rr}$ given by equation (\ref{eq:A}) the energy-density
turns out to be:

\begin{equation} \rho(r) = \frac{1}{8 \pi r^2} (1-\theta(r-r_h))
\end{equation}

Within the holostar's interior the mass-energy density follows an
inverse square law. Outside of the membrane, i.e. for $r > r_h$,
it is identical zero. Note, that $r_h$ must not necessarily be
finite.

In the following discussion the argument $(r
- r_h)$ of the $\theta$- and $\delta$-distributions will be omitted.

The radial and tangential pressures also follow from the metric:

\begin{equation} \label{eq:PrBH}
P_r = -\rho = -\frac{1}{8 \pi r^2} (1-\theta)
\end{equation}

\begin{equation}
P_\theta = P_\varphi = \frac{1}{16 \pi r_h} \delta
\end{equation}

$P_r$ is the radial pressure. It is equal in magnitude but
opposite in sign to the mass-density. $P_\theta$ denotes the
tangential pressure, which is zero everywhere, except for a
$\delta$-functional at the holostar's surface. The
"stress-energy-content" of the two principal tangential pressure
components in the membrane is equal to the gravitating mass $M$ of
the holostar.

In order to determine the principal pressures from the metric, the
time-coefficient of the metric $g_{tt}$ must be known. For the
holostar equation of state with $P_r = - \rho$ we have $g_{tt} =
1/g_{rr}$. Other equations of state lead to different
time-coefficients, and therefore different principal pressures.

Neither the particular form of the time-coefficient of the metric,
nor the particular form of the principal pressures are important
for the main results derived in this paper, which are based on
equilibrium thermodynamics, where time evolution is irrelevant (as
long as the relevant time scale is long enough, that thermal
equilibrium can be attained). The essential assumptions are:
\begin{itemize}
\item spherical symmetry \item a radial
metric coefficient $g_{rr} = r/r_0$ \item a total energy density
$\rho = 1 / (8 \pi r^2)$ \item microscopic statistical
thermodynamics of an ideal gas of ultra-relativistic fermions and
bosons (in the context of the grand-canonical ensemble)
\end{itemize}

If the validity of Einstein's field equations with zero
cosmological constant is assumed, conditions two and three are
interchangeable.

Throughout this paper I will frequently use the term holographic
solution, or holostar, to refer to an object with the above stated
properties. The reader should keep in mind, though, that the
holographic solution is just a special case of a solution with a
matter-density $\rho = 1 /(8\pi r^2)$. The results derived in this
paper refer to any solution with the above properties.

In the following sections I assume that $r_0^2$ is nearly
constant, i.e. more or less independent of the size of the
holostar and comparable to the Planck area $A_{Pl} = \hbar$:

\begin{equation} \label{eq:r0}
{r_0}^2 = \beta {r_{Pl}}^2 = \beta \hbar
\end{equation}

This assumption will be justified later.

\section{\label{sec:fermion:simple}A simple derivation of the Hawking temperature and entropy}

The interior metric of the holostar solution is well behaved and
the interior matter-density is non-zero. The solution is static:
The matter appears to exert a radial pressure preventing further
collapse to a point singularity. This can be best seen in the
string picture \cite{petri/string}. However, the solution gives no
direct indication with respect to the state of the interior matter
and the origin of the pressure.

In this section I will discuss a very simple model for the
interior matter state of the holostar, which is able to explain
many phenomena attributed to black holes. Let us assume that the
interior matter distribution is dominated by ultra-relativistic
weakly interacting fermions and the pressure is produced by the
exclusion principle. Due to spherical symmetry the mean momentum
of the fermions $p(r)$ and their number density per proper volume
$f n(r)$  will only depend on the radial distance coordinate $r$.
$f$ denotes the effective number of degrees of freedom of the
fermions. For ultra-relativistic fermions the local energy-density
will be given by the product of the number density of the fermions
and their mean momentum. This energy density must be equal to the
interior mass-energy density of the holostar:

\begin{equation} \label{eq:rho_fnp}
\rho = p(r) f n(r) = \frac{1}{8 \pi r^2}
\end{equation}

If the fermions interact only weakly, their mean momenta can be
estimated by the exclusion principle:

\begin{equation} \label{eq:p3n}
p(r)^3 \frac{1}{n(r)} = (2 \pi \hbar)^3
\end{equation}

These two equations can be solved for $p(r)$ and $n(r)$:

\begin{equation} \label{eq:plocr}
p(r) = \frac{\hbar^{\frac{3}{4}}
\pi^{\frac{1}{2}}}{f^{\frac{1}{4}}} \frac{1}{r^{\frac{1}{2}}}
\end{equation}

\begin{equation} \label{eq:nlocr}
f n(r) = \frac{f^{\frac{1}{4}}}{\hbar^{\frac{3}{4}} 8
\pi^{\frac{3}{2}}} \frac{1}{r^{\frac{3}{2}}}
\end{equation}

The mean momenta of the fermions within the holostar fall off from
the center as $1/r^{1/2}$ and the number density per proper volume
with $1/r^{3/2}$. Similar dependencies, however without definite
factors, have already been found in \cite{petri/hol} by analyzing
the geodesic motion of the interior massless particles in the
holostar-metric. It is remarkable, that equilibrium thermodynamics
combined with the uncertainty principle gives the same results as
the geodesic equations of motion. This is not altogether
unexpected. In \cite{Jacobson} it has been shown, that the field
equations of general relativity follow from thermodynamics and the
Bekenstein entropy bound \cite{Bekenstein/81}.

The momentum of the fermions at a Planck-distance $r = r_{Pl} =
\sqrt{\hbar}$ from the center of the holostar is of the order of
the Planck-energy $E_{Pl}=\sqrt{\hbar}$. It is also interesting to
note, that for both quantities $p(r)$ and $n(r)$ the number of
degrees of freedom $f$ can be absorbed in the radial coordinate
value $r \rightarrow \sqrt{f} \, r$, so that $p$ and $n$
effectively depend on $\sqrt{f} \, r$. We will see later that the
square root of $f$ plays an important role in the scaling of the
fundamental length parameter $r_0$.

From (\ref{eq:plocr}) one can derive the following momentum-area
law for holostars, which resembles the Stefan-Boltzmann law for
radiation from a black body:

\begin{equation} \label{eq:p4r2}
p(r)^4 r^2 f = \hbar^3 \pi^2
\end{equation}

Note that this law not only refers to the holostar's surface ($r
= r_h$) but is valid for any concentric spherical surface of
radius $r$ within the holostar. Therefore it is reasonable to
assume that the holostar has a well defined interior temperature
$T(r)$ proportional to the mean momentum $p(r)$:

\begin{equation} \label{eq:pT}
p(r) = \sigma T(r)
\end{equation}

$\sigma$ is a constant factor. We will see later, that it is related to
the entropy per particle.

The local surface temperature of the holostar is given by:

\begin{equation}
T(r_h) = \frac{p(r_h)}{\sigma} = \frac{\hbar^{\frac{3}{4}}
\pi^{\frac{1}{2}}}{\sigma f^{\frac{1}{4}}} \frac{1}{\sqrt{r_h}}
\end{equation}

The surface redshift $z$ is given by:

\begin{equation} \label{eq:z}
z = \frac{1}{\sqrt{g_{tt}(r_h)}} = \sqrt{g_{rr}(r_h)}=
\sqrt{\frac{r_h}{r_0}}
\end{equation}

where $g_{tt}(\infty) = 1$ is assumed.

The local surface temperature can be compared to the Hawking
temperature of a black hole. The Hawking temperature is measured
at infinity. Therefore the red-shift of the radiation emitted from
the holostar's surface with respect to an observer at spatial
infinity has to be taken into account, by dividing the local
temperature at the surface by the gravitational red shift factor
$z$. With $g_{rr}(r_h) = {r_h}^{1/2} ({\beta \hbar})^{-1/4}$  we
find:

\begin{equation} \label{eq:T_inf}
T_{\infty}= \frac{T(r_h)} {\sqrt{g_{rr}(r_h)}}=
\frac{\pi^{\frac{1}{2}}}{\sigma}
\left(\frac{\beta}{f}\right)^{\frac{1}{4}} \frac{\hbar}{r_h}
\end{equation}

The surface-temperature measured at infinity has the same
dependence on the gravitational radius $r_h$ as the Hawking
temperature, which is given by:

\begin{equation} \label{eq:T_Hawking}
T_{H} = \frac{\hbar}{4 \pi r_h} = \frac{\hbar}{8 \pi M}
\end{equation}

Up to a possibly different constant factor the Hawking temperature
of a spherically symmetric black hole and the respective
temperature of the holostar at infinity are equal.

As the Hawking temperature of a black hole only depends on the
properties of the exterior space-time, and the exterior
space-times of a black hole and the holostar are equal (up to a
small Planck-sized region outside the horizon), it is reasonable
to assume that the Hawking temperature should be the true
temperature of a holostar measured at spatial infinity. With this
assumption, the constant $\sigma$ can be determined by setting the
temperatures of equations (\ref{eq:T_inf}) and
(\ref{eq:T_Hawking}) equal:

\begin{equation} \label{eq:s} \label{eq:beta}
\sigma = \left(\frac{\beta}{f}\right)^{\frac{1}{4}} 4 \pi^{\frac{3}{2}}
\end{equation}

The total number of fermions within the holostar is given by the
proper integral over the number-density:

\begin{equation} \label{eq:Nint}
N = \int{f n(r) dV}
\end{equation}

$dV$ is the proper volume element, which can be read off from the
metric:

\begin{equation} \label{eq:dV}
dV = 4 \pi r^2 \sqrt{g_{rr}} dr = 4 \pi r^{\frac{5}{2}}
\left({\beta \hbar} \right)^{-\frac{1}{4}} dr
\end{equation}

Integration over the total interior volume of the holostar gives:

\begin{equation} \label{eq:Nclassic}
N = \left(\frac{f}{\beta}\right)^{\frac{1}{4}} \frac{1}{4
\pi^{\frac{3}{2}}} \frac{\pi r_h^2}{\hbar} = \frac{1}{\sigma}
\frac{A}{4 \hbar} = \frac{S_{BH}}{\sigma}
\end{equation}

$S_{BH}$ is the Bekenstein-Hawking entropy for a spherically
symmetric black hole with horizon surface area $A$.

Therefore the number of fermions within the holostar is
proportional to its surface area and thus proportional to the
Hawking entropy. This result is very much in agreement with the
holographic principle \cite{Hooft/hol, Susskind/hol}, giving it
quite a new and radical interpretation: The degrees of freedom of
a highly relativistic self-gravitating object don't only "live on
the surface", the object contains a definite number of particles
and their total number is proportional to the object's surface
area, measured in units of the Planck-area, $A_{Pl} = \hbar$. This
result is an immediate consequence of the interior metric $g_{rr}
\propto r$, the energy-momentum relation for relativistic
particles $E = p$ and the exclusion principle. It can be easily
shown, that for any other spherically symmetric metric, for
example $g_{rr} \propto r^n$, the number of interior (fermionic)
particles is not proportional to the boundary area.

From equation (\ref{eq:Nclassic}) we can see that $\sigma$ is the
entropy per particle. This allows a rough estimate of $\beta$: The
entropy of an ultra-relativistic particle should be of order unity
($\sigma \approx 3-4$). The degrees of freedom in the Standard
Model of particle physics - with the usual counting rule,
weighting the fermionic degrees of freedom with $7/8$ - amount to
$f \approx 100$. Supersymmetry essentially doubles this number. It
is expected, that a unified theory will not vastly exceed this
number. For $\sigma = 3$ and $f = 256$ we find $4 \pi \beta \simeq
1.06$. This justifies the assumption, that the fundamental length
parameter $r_0$ should be of order Planck-length.

By help of equation (\ref{eq:s}) the local temperature can be
expressed in terms of $\beta$ alone:

\begin{equation} \label{eq:Tlocal}
T(r) = \frac{\hbar^{\frac{3}{4}}}{4 \pi \beta^{\frac{1}{4}}}
\frac{1}{r^{\frac{1}{2}}} = \frac{1}{4 \pi} \frac{\hbar}{(r_0
r)^{\frac{1}{2}}}
\end{equation}

Note that $\beta$ depends explicitly on the (effective) number of
degrees of freedom $f$ of the ultra-relativistic particles within
the holostar via equation (\ref{eq:beta}). At the center of the
holostar all the fermion momenta are comparable to the Planck
energy, as can be seen from equation (\ref{eq:Tlocal}). All
fermions of the Standard Model of particle physics will be
ultra-relativistic. Quite likely there will be other fundamental
particles of a grand unified theory (GUT), as well as other
entities such as strings and branes. Thus, close to the holostar's
center the number of ultra-relativistic degrees of freedom will be
at its maximum and $\beta$ will be close to unity. The farther one
is distanced from the center, the lower the local temperature
gets. At $r \approx 10^6 km$ the electrons will become
non-relativistic. The only particles of the Standard Model that
remain relativistic at larger radial positions will be the
neutrinos. If all neutrinos are massive, the mass of the lightest
neutrino will define a characteristic radius of the holostar,
beyond which there are no relativistic fermions contributing to
the holostar's internal pressure. If at least one of the neutrinos
is massless, there will be no limit to the spatial extension of a
holostar.

Note, that the radial coordinate position at which the holostar's
interior radiation temperature is equal to the temperature of the
cosmic microwave background radiation, $T_{CMBR} = 2.725 \, K$,
corresponds to roughly $r \approx 10^{28} m\approx10^{12} \, ly$,
i.e. quite close to the radius of the observable universe. This is
just one of several coincidences, which point to the very real
possibility, that the holostar or a variant thereof actually might
serve as an alternative, beautifully simple model for the
universe. For a more detailed discussion including some definite
cosmological predictions, which are all experimentally verified
within an error of maximally 15 \% see \cite{petri/hol}.

Whenever the temperature within the holostar becomes comparable to
the mass of a particular fermion species, a phase transition is
expected to take place at the respective $r$-position. Such a
transition will lower the effective value of $f$, as one of the
particles "freezes" out. Whenever $f$ changes, either $\sigma$ or
$\beta$ must adjust due to equation (\ref{eq:beta}). The question
is, whether $\sigma$ or $\beta$ (or both) will change. Presumably
$\sigma$ will at least approximately retain a constant value: The
entropy per ultra-relativistic fermion, as well as the mean
particle momentum per temperature, appears to be a local property
which should not depend on the (effective) number of degrees of
freedom of the particles at a particular $r$-position.

Under the assumption that $\sigma$ is nearly constant, the ratio of
$\beta / f$ must be nearly constant as well, as can be seen
from equation (\ref{eq:s}). Whenever $f$ changes, $\beta$ will
adjust accordingly. Lowering the effective number of degrees of
freedom leads to a flattening of the temperature-curve, as heat
(and entropy) is transferred to the remaining ultra-relativistic
particles. At any radial position of a phase transition, where a
fermion becomes non-relativistic and annihilates with its
anti-particle, the temperature is expected to deviate from the
expression $T \propto 1/\sqrt{r}$. This is quite similar to what
is believed to have happened in the very early universe, when the
temperature fell below the electron-mass threshold and the
subsequent annihilation of electron/positron pairs heated up the
photon gas, keeping the temperature of the expanding universe
nearly constant until all positrons were destroyed.

If the "freeze-out" happens without significant heat and entropy
transfer to the remaining gas of ultra-relativistic particles,
such as when the particle that "freezes" out has an appreciable
non-zero chemical potential, the effective value of $f$ will
remain nearly constant, which would imply that $\beta$ be nearly
constant as well. In this case $\beta$ as well as $f$ would be
nearly constant universal quantities. There is evidence that this
might actually be the case.\footnote{See \cite{petri/charge,
petri/hol, petri/thermo}.}

\section{\label{sec:fermionBosonGas}Thermodynamics of an ultra-relativistic fermion and boson gas}

In this section I will discuss a somewhat more sophisticated model
for the thermodynamic properties of the holostar.

As has been demonstrated in the previous section, if the holostar
contains at least one fermionic species, its properties very much
resemble the Schwarzschild vacuum black hole solution, when viewed
from the outside: Due to Birkhoff's theorem the external
gravitational field cannot be distinguished from that of a
Schwarzschild black hole. Its temperature measured at infinity is
proportional to the Hawking temperature.

Due to its non-zero surface-temperature and entropy the holostar
will gradually lose particles by emission from its surface. The
(exterior) time scale of this process will be comparable to the
Hawking evaporation time scale $\propto r_h^3$ (see for example
\cite{petri/hol}). The (exterior) time for a photon to travel
radially through the holostar is proportional to $r_h^2$.
Therefore even comparatively small holostars are expected to have
an evaporation time several orders of magnitude longer than their
interior relaxation time.

This allows us, with the possible exception of near Planck-size
holostars, to consider any spherical thin shell within the
holostar's interior to be in thermal equilibrium with its
surroundings. Each shell can exchange particles, energy and
entropy with adjacent shells on a time scale much shorter than the
life-time of the holostar. Under these assumptions the
thermodynamic parameters within each shell can be calculated via
the grand canonical ensemble.

We mentally partition the holostar into a collection of thin
spherical shells. The temperature scales as $1/\sqrt{r}$ and thus
varies very slowly with $r$. For the chemical potential(s) let us
assume a slowly varying function with $r$ as well. This assumption
will be justified later. Under these circumstances the thickness
of each shell $\delta r$ can be chosen such, that it is large
enough to be considered macroscopic, and at the same time small
enough, so that the temperature, pressure and chemical
potential(s) are effectively constant within the shell.

An accurate thermodynamic description has to take into account a
possible potential energy of position. For the holostar a
significant simplification arises from the fact, that the
effective potential $V_{eff}(r)$ for the motion of massless, i.e.
ultra-relativistic, particles is nearly constant. The equations of
motion for ultra-relativistic particles within the holostar's
interior were given by \cite{petri/hol}:

\begin{equation} \label{eq:EqMotion:r}
V_{eff}(r) = \frac{r_i^3}{r^3}
\end{equation}

and

\begin{equation} \label{eq:EqMotion:t}
\beta_\perp^2(r) = \frac{r_i^3}{r^3}
\end{equation}

$\beta_r(r)$ is the radial velocity of a photon, expressed as
fraction to the local velocity of light in the (purely) radial
direction. $\beta_\perp(r)$ is the tangential velocity of the
photon, expressed as a fraction to the local velocity of light in
the (purely) tangential direction. $r_i$ is the turning point of
the motion. For pure radial motion $r_i=0$.

For pure radial motion the effective potential is constant with
$V_{eff}(r) = 0$. In the case of angular motion ($r_i \neq 0$) the
effective potential approaches zero with $1/r^3$, i.e. becomes
nearly zero very rapidly, whenever $r$ is greater than a few
$r_i$. Therefore, to a very good approximation we can regard the
ultra-relativistic particles to move freely within each shell.
Their total energy will only depend on the relativistic
energy-momentum relation, not on the radial position.

With these preliminaries the grand canonical potential $\delta J$
of a small spherical shell of thickness $\delta r$ for a gas of
relativistic fermions at radial position $r$ will be given by:

$$
\delta J(r) = - T(r) \frac{f}{(2 \pi \hbar)^3} \delta V \int \int
\int d^3p \ln{(1+e^{-\frac{p- \mu(r)}{T(r)}})}
$$

\begin{equation} \label{eq:J0}
 = - T^4 \delta V \frac{f}{2 \pi^2 \hbar^3} \int_{0}^{\infty}{z^2
\ln{(1+e^{-z+\frac{\mu}{T}})} dz}
\end{equation}

$z = p/T(r)$ is a dimensionless integration variable. $\mu(r)$ is
the chemical potential at radial coordinate position $r$. $T(r)$
is the local temperature at this position.

Note that even when the radial coordinate extension $\delta r$ of
the shell is small, the {\em proper} radial extension $\delta l =
(r/r_0)^{1/2} \delta r$ of the shell will become quite large
because of the large value of the radial metric coefficient in the
holostar's outer regions.

Knowing the results presented at the end of this section it is not
difficult to show that - with the exception of the central region
- it is possible to choose the radial extension of the shell such
that the number of particles within the shell $N$ is macroscopic
and at the same time $T(r)$ and $\mu(r)$ are constant to a very
good approximation within the shell. The proper volume of the
shell $\delta V$ is given by the volume element of equation
(\ref{eq:dV}).

The ratio of chemical potential $\mu$ to local temperature $T$ is
assumed to be a very slowly varying function of $r$. In fact, we
will see later that this ratio is virtually independent of $r$.
The ratio $\mu/T$ will be denoted by $u$, keeping in mind that $u$
might depend on $r$:

$$u = \frac{\mu(r)}{T(r)}$$

The integral in equation (\ref{eq:J0}) can be transformed to the
following integral by a partial integration:

\begin{equation} \label{eq:J}
\delta J(r) = - T^4 \delta V \frac{f}{2 \pi^2 \hbar^3} \frac{1}{3}
\int_0^\infty{z^3 n_F(z,u)dz}
\end{equation}

where $n_F$ is the mean occupancy number of the fermions:

\begin{equation} \label{eq:nF}
n_F(z, u) = \frac{1}{e^{z-u}+1} = \frac{1}{e^{\frac{p-\mu}{T}}+1}
\end{equation}

Knowing the grand canonical potential $\delta J$ the entropy
within the shell can be calculated:

\begin{equation} \label{eq:SF}
\delta S(r) = -\frac{\partial (\delta J)}{\partial T} = \frac{f}{2
\pi^2 \hbar^3} T^3 \delta V \left(\frac{4}{3} Z_{F,3}(u) - u
Z_{F,2}(u)\right)
\end{equation}

By $Z_{F,n}$ the following integrals are denoted:

\begin{equation} \label{eq:ZF}
Z_{F,n}(u) = \int_0^\infty{z^n n_F(z,u)dz}
\end{equation}

Such integrals commonly occur in the evaluation of
Feynman-integrals in QFT and can be evaluated by the
poly-logarithmic function $Li_n(z)$:

\begin{equation} \label{eq:ZF:Li}
Z_{F,n}(u) = -\Gamma(n+1) \, Li_{n+1}(-e^u)
\end{equation}

For the derivation of the entropy the following identity has been
used, which is easy to derive from the power-expansion of
$Li_n(z)$.

\begin{equation}
\frac{\partial {Z_{F,3}(u)}}{\partial x} = 3 Z_{F,2}(u)
\frac{\partial u}{\partial x}
\end{equation}

The pressure in the shell is given by:

\begin{equation} \label{eq:PF}
P(r) = -\frac{\partial (\delta J)}{\partial (\delta V)} =
\frac{f}{2 \pi^2 \hbar^3} T^4 \frac{Z_{F,3}(u)}{3}
\end{equation}

The total energy in the shell can be calculated from the grand
canonical potential via:

\begin{equation} \label{eq:EF}
\delta E(r) = \delta J -\left(T \frac{\partial}{\partial T} + \mu
\frac{\partial}{\partial \mu}\right)\delta J = \frac{f}{2 \pi^2
\hbar^3} T^4 \delta V Z_{F,3}(u)
\end{equation}

The total number of particles within the shell is given by:

\begin{equation} \label{eq:NF}
\delta N(r) = -\frac{\partial (\delta J)}{\partial \mu}=\frac{f}{2
\pi^2 \hbar^3} T^3 \delta V Z_{F,2}(u)
\end{equation}

The total energy per fermion within the shell is proportional to
$T$, as can be seen by combining equations (\ref{eq:EF},
\ref{eq:NF}):

\begin{equation} \label{eq:EF_NF}
\epsilon = \frac{\delta E}{\delta N} =
\frac{Z_{F,3}(u)}{Z_{F,2}(u)} \, \, T(r)
\end{equation}

$\epsilon$ only depends indirectly on $r$ via $u$. We will see
later that $u$ is essentially independent of $r$, so that the mean
energy per particle is proportional to the temperature with nearly
the same constant of proportionality at any radial position $r$.

The entropy per particle within the shell can be read off from
equations (\ref{eq:SF}, \ref{eq:NF}):

\begin{equation} \label{eq:SF_NF}
\sigma = \frac{\delta S}{\delta N} = \frac{4}{3}
\frac{Z_{F,3}(u)}{Z_{F,2}(u)} - u
\end{equation}

Again, $\sigma$ only depends on $r$ via $u$.

The calculations so far have been carried through for fermions. It
is likely, that the holostar will also contain bosons in thermal
equilibrium with the fermions. The equations for an
ultra-relativistic boson gas are quite similar to the above
equations for a fermion gas. We have to replace:

\begin{equation} \label{eq:Replace:nB}
n_F(z,u) \rightarrow n_B(z,u) = \frac{1}{e^{z-u}-1}
\end{equation}

\begin{equation} \label{eq:Replace:zB}
Z_{F,n} \rightarrow Z_{B,n} = \int_0^\infty{z^n n_B(z,u) dz}
\end{equation}

\begin{equation} \label{eq:polylog:b}
Z_{B,n} = \Gamma(n+1) \, Li_{n+1}(e^u)
\end{equation}

Let us assume that the fermion and boson gases are only weakly
interacting. In such a case the extrinsic quantities, such as
energy and entropy, can be simply summed up. The same applies for
the partial pressures.

The number of degrees of freedom of fermions and bosons can
differ. The fermionic degrees of freedom will be denoted by $f_F$,
the bosonic degrees of freedom by $f_B$. In general, the different
particle species will have different values for the chemical
potentials. There are some restraints. Bosons cannot have a
positive chemical potential, as $Z_{B,n}(u)$ is a complex number
for positive $u$. Photons and gravitons, in fact all massless
gauge-bosons, have a chemical potential of zero, as they can be
created and destroyed without being restrained by a
particle-number conservation law.

We are however talking of a gas of ultra-relativistic particles.
In this case particle-antiparticle pair production will take place
abundantly, so that we also have to consider the antiparticles.
The chemical potentials of particle and anti-particle add up to
zero: $\mu + \overline{\mu} = 0$. As bosons cannot have a positive
chemical potential, the chemical potential of any
ultra-relativistic bosonic species must be zero, i.e. $\mu_B =
\overline{\mu_B} = 0$, whenever the energy is high enough to
create boson/anti-boson pairs. This restriction does not apply to
the fermions, which can have a non-zero chemical potential at
ultra-relativistic energies, as both signs of the chemical
potential are allowed. So for ultra-relativistic fermions we can
fulfill the relation $\mu_F + \overline{\mu_F} = 0$ with non-zero
$\mu_F$.

It is convenient to use the ratio of the chemical potential to the
temperature $u = \mu / T$ as the relevant parameter instead of the
chemical potential itself . If the number of degrees of freedom of
fermions and bosons respectively, i.e. $f_F$ and $f_B$ is known,
there are only two undetermined parameters in the model, $u_F$ and
$\beta$. In order to determine $u_F$ and $\beta$ one needs two
independent relations. These can be obtained by comparing the
holostar temperature and entropy to the Hawking temperature and
entropies respectively.

Alternatively $u_F$ can be determined without reference to the
Hawking temperature law, solely by a thermodynamic argument. It is
also possible to determine $\beta$ by a theoretical argument as
proposed in \cite{petri/charge}.

The thermodynamic energy of a shell consisting of an
ultra-relativistic ideal fermion and boson gas is given by:

\begin{equation} \label{eq:E_th}
\delta E_{th} = \frac{F_E} {2 \pi^2 \hbar^3}  \delta V T^4
\end{equation}

with

\begin{equation}
F_E(u_F) = f_F ( Z_{F,3}(u_F) + Z_{F,3}(-u_F)) + 2 f_B Z_{B,3}(0)
\end{equation}

with the identities of the polylog-function and with $Z_{B,3}(0) =
\pi^4 / 15$ one can express $F_E$ as a quadratic function of
$u_F^2/\pi^2$ \cite{petri/asym}:

\begin{equation} \label{eq:FE:u2}
F_E(u_F) = 2 f_F \frac{\pi^4}{15} \left(\frac{15}{8}\left(1 +
\frac{\pi^2}{u_F^2} \right)^2 + \frac{f_B}{f_F} -1 \right)
\end{equation}

We take the convention here, that $f_F$ and $f_B$ denote the
degrees of freedom of one particle species, including particle and
antiparticle. With this convention a photon gas ($g=2$) is
described by $f_B = 1$ (There are two photon degrees of freedom
and the photon is its own anti-particle). All other particle
characteristics, such as helicities, are counted extra. The total
number of the degrees of freedom in the gas, i.e. counting
particles and anti-particles separately, will be given by

\begin{equation}
f = 2(f_F + f_B)
\end{equation}

The total energy of the holostar solution is given by the
proper integral over the mass density. The proper energy of the
shell therefore is:

\begin{equation} \label{eq:E_BH}
\delta E_{BH} = \rho \delta V = \frac{\delta V}{8 \pi r^2} =
\frac{1}{2} (\beta \hbar)^{-\frac{1}{4}} r^{\frac{1}{2}} \delta r
\end{equation}

Setting the two energies equal gives the local temperature within
the holostar:

\begin{equation} \label{eq:TlocTherm}
T^4 = \frac{\pi \hbar^3}{4 F_E r^2 }
\end{equation}

Thus we recover the $1/\sqrt{r}$-dependence of the local
temperature, at least if $F_E$ is constant.

$F_E$ is a function of $f_F$, $f_B$ and $u_F$. We will see later,
that $u_F$ only depends on the ratio of $f_F$ and $f_B$. Therefore
in any range of $r$-values where the number of degrees of freedom
of the ultra-relativistic particles (or rather their ratio)
doesn't change, the local temperature as determined by equation
(\ref{eq:TlocTherm}) will not deviate from an inverse square root
law.

If the temperature of equation (\ref{eq:TlocTherm}) is inserted
into equation (\ref{eq:PF}), the thermodynamic pressure is derived
as follows:

$$P(r) = \frac{1}{24 \pi r^2} = \frac{\rho}{3}$$

This is the equation of state for an ultra-relativistic gas, as
expected.

\subsection{Comparing the holostar's thermodynamic temperature
and entropy to the Hawking result}

By inserting the temperature derived in equation
(\ref{eq:TlocTherm}) into equation (\ref{eq:SF}) we get the
following expression for the thermodynamic entropy within the
shell:

\begin{equation} \label{eq:Sshell}
\delta S(r) = \left(\frac{F_E}{4 \pi \beta}\right)^{\frac{1}{4}}
 \frac{F_S}{F_E} \frac{r \delta r}{\hbar}
\end{equation}

with

\begin{equation} \label{eq:FS}
F_S(u_F) = f_F \left( \frac{4}{3}  \{Z_{F,3}(u_F)\} - u_F
[Z_{F,2}(u_F)] \right) + 2 f_B \frac{4}{3}  \left( Z_{B,3}(0)
\right)
\end{equation}

We have used commutator [] and anti-commutator \{\} notation in
order to render the above relation somewhat more compact.

Using the identities for the polylog function it is possible to
express the above relation as a quadratic function of the variable
$u_F^2 / \pi^2$.

\begin{equation}
F_S = \frac{4}{3} F_E(u_F) - f_F \frac{\pi^4}{3} \frac{u_F^2}{\pi^2} \left(1 + \frac{u_F^2}{\pi^2} \right)
\end{equation}

with $F_E$ is given by equation (\ref{eq:FE:u2})

By comparing the temperature (\ref{eq:TlocTherm}) and the entropy
(\ref{eq:Sshell}) of the holostar solution derived in the context
of our simple model to the Hawking entropy and temperature, two
important relations involving the two unknown parameters of the
model $u_F$ and $\beta$ can be obtained.

We have already seen in section \ref{sec:fermion:simple} that the
holostar's temperature at infinity is proportional to the Hawking
temperature. As can be seen from equation (\ref{eq:TlocTherm})
this general result remains unchanged in the more sophisticated
thermodynamic analysis, as long as the quantity $F_E(u_F, f_F,
f_B)$ can be considered to be nearly constant. In order to
determine $F_E$ we can set the temperature at the holostar's
surface equal to the blue shifted Hawking temperature at the
holostar's surface, which can be obtained by multiplying the
Hawking temperature (at infinity) with the red-shift factor $z$ of
the surface given in Eq. (\ref{eq:z}). We find:

\begin{equation} \label{eq:TBH4}
T^4 = {T_{BH}}^4 \,  z^4 = \frac{\hbar^4}{2^8 \pi^4 {r_h}^4} \cdot
\frac{r_h^2}{\beta \hbar} = \frac{\hbar^3}{2^8 \pi^4 \beta
{r_h}^2}
\end{equation}

Comparing this to equation (\ref{eq:TlocTherm}) we find:

\begin{equation} \label{eq:FE:beta}
\frac{F_E}{4 \pi \beta} = (2 \pi)^4
\end{equation}

This is an important result. It relates the fundamental area $4
\pi r_0^2 = 4 \pi \beta \hbar$ to the thermodynamic parameters of
the system, i.e. the number of degrees of freedom and the chemical
potential of the fermions.

Another important relation  is the ratio $F_S/F_E$ in the interior
holostar space-time, which can be obtained by comparing the
Hawking entropy of a black hole with thermodynamic entropy of the
holostar's interior constituent matter.

The entropy of the holostar can be calculated by integrating
equation (\ref{eq:Sshell}). We will assume that $F_E / \beta =
const$, as follows from equation (\ref{eq:FE:beta}), and that
$F_S/F_E = const$, which will be justified shortly. If this is the
case, the integral can be performed easily:

\begin{equation}
S = \int_0^{r_h}{\delta S(r) dV} = \left(\frac{F_E}{4 \pi
\beta}\right)^{\frac{1}{4}} \frac{1}{2 \pi}
 \frac{F_S}{F_E} \frac{A}{4 \hbar}
\end{equation}

with

$$A = 4 \pi {r_h}^2$$

Setting this equal to the Hawking entropy, $S_{BH} = A/(4 \hbar)$,
and using equation (\ref{eq:FE:beta}) we find the important
result:

\begin{equation} \label{eq:FS:FE}
\frac{F_S}{F_E} = 1
\end{equation}

Writing out the above equation we get:

\begin{equation} \label{eq:ImplicituF}
\frac{f_F\left(\frac{4}{3} \{Z_{F,3}(u_F)\} - u_F
[Z_{F,2}(u_F)]\right)+ 2 f_B\left(\frac{4}{3}
Z_{B,3}(0)\right)}{f_F \{Z_{F,3}(u_F)\} + 2 f_B Z_{B,3}(0)} = 1
\end{equation}

Using the identities for the polylog function one can reduce the
above equation to a very simple quadratic equation in the variable
$u_F^2 / \pi^2$:

\begin{equation} \label{eq:uF:expl}
\left(1 + \frac{u_F^2}{\pi^2} \right) \left(1 - 3
\frac{u_F^2}{\pi^2} \right) + \frac{8}{15} \left(\frac{f_B}{f_F}
-1\right) = 0
\end{equation}

The important message is, that whenever the bosonic and fermionic
degrees of freedom - or rather their ratio $f_B/f_F$ - is known,
$u_F$ can be calculated. Knowing $u_F$, $\beta$ can be determined
via (\ref{eq:FE:beta}). Thus the two relations (\ref{eq:FE:beta},
\ref{eq:FS:FE}) allow us to determine all free parameters of the
model, whenever the number of particle degrees of freedom, $f_F$
and $f_B$ are known.

\subsection{An alternative derivation of the relation $F_S/F_E = 1$}

Before discussing the specifics of the thermodynamic model, I
would like to point out another derivation of equation
(\ref{eq:FS:FE}), which does not depend on the Hawking result.
This alternative derivation only depends on the following
fundamental thermodynamic relation

\begin{equation} \label{eq:SET}
\frac{\delta S}{\delta E} T = 1
\end{equation}

and on the fact, that the holostar's interior matter state is
completely rigid, i.e. the interior matter state at any particular
radial position depends only on $r$, but not on the overall size
of the holostar.

Consider a process, where an infinitesimally small spherical shell
of matter is added to the outer surface of the holostar. This
process doesn't affect the inner matter of the holostar, as the
interior matter-state of the holostar at a given radial
coordinated position $r$ does not depend in any way on the size of
the holostar or on any other global quantity. Therefore, when
adding a new layer of matter we don't have to consider any
interaction, such as heat-, energy- or entropy-transfer between
the newly added matter layer and the interior matter. It is an
adiabatic process, for which we can calculate the entropy-change
of the whole system via equation (\ref{eq:SET}). Let $r$ be the
radial position of the holostar's surface. The entropy of the
newly added shell is given by equation (\ref{eq:Sshell}), its
energy by equation (\ref{eq:E_BH}), and its temperature by
equation (\ref{eq:TlocTherm}). One finds that the thermodynamic
relation (\ref{eq:SET}) is only fulfilled, when $F_S = F_E$. We
have derived equation (\ref{eq:FS:FE}) only from thermodynamics.

\subsection{A closed formula for $u_F$ and some special cases}

The chemical potential per temperature $u_F$ can be determined by
finding the root of equation (\ref{eq:uF:expl}). The value of
$u_F$ depends only on the ratio of fermionic to bosonic degrees of
freedom.\footnote{and on the constant ratio $F_S/F_E$, which has
been shown to be unity for the interior holostar solution}. Let us
denote the ratio of the degrees of freedom by

\begin{equation}
r_f = \frac{f_B}{f_F}
\end{equation}

Then $u_F$ is given by:

\begin{equation} \label{eq:u2}
\frac{u_F^2}{\pi^2} = \frac{2}{3} \sqrt{1 + \frac{2}{5} (r_f-1)} -
\frac{1}{3}
\end{equation}

For $r_f=0$ (only fermions) we find the following result:

\begin{equation} \label{eq:uF:rf=0}
u_F = \pi \, \sqrt{\sqrt{\frac{4}{15}} - \frac{1}{3}}  = 1.34416
\end{equation}

For $r_f=1$ (equal number of fermions and bosons) we get:

\begin{equation} \label{eq:uF;rf=1}
u_F = \frac{\pi}{\sqrt{3}} = 1.8138
\end{equation}

From equation (\ref{eq:u2}) one can see that $u_F$ is a
monotonically increasing function of $r_f$. It attains its minimum
value, when there are no bosonic degrees of freedom, i.e. $f_B =
r_f = 0$. When the bosonic degrees of freedom vastly exceed the
fermionic degrees of freedom $u_F$ can - in principle - attain
high values. For large $r_f$ we have $u_F \propto (r_f-1)^{1/4}$.
For all practical purposes one can assume that the number of
bosonic degrees of freedom is not very much higher than the number
of fermionic degrees of freedom. This places $u_F$ in the range
$1.34 < u_F < 3$.

It is important to notice, that equation (\ref{eq:ImplicituF})
only has a solution when the number of fermionic degrees of
freedom, $f_F$, is non-zero, whereas $f_B$ can take arbitrary
values for any non-zero $f_F$. Therefore at least one fermionic
(massless) particle species with a non-vanishing chemical
potential proportional to the local radiation temperature is
necessary.

\subsection{Thermodynamic relations, which are independent from the Hawking formula}

If $u_F$ is known, all thermodynamic quantities of the model, such
as $F_E(u_F)$ and $F_N(u_F)$ etc. can be evaluated. Note that in
order to determine $u_F$ we only needed the relation $F_E = F_S$,
whose derivation didn't require the Hawking temperature/entropy
relation. Yet in order to fix $\beta$ via equation
(\ref{eq:FE:beta}) we had to compare the holostar's temperature
(or entropy) to the Hawking-result. Therefore the particular
relation between $\beta$ and $F_E$ derived in equation
(\ref{eq:FE:beta}) is tied to the the validity of the Hawking
temperature formula.

Although there is no doubt that the Hawking temperature of a large
black hole must be inverse proportional to its mass\footnote{This
already follows from the Bekenstein-argument, that the entropy of
a black hole should be proportional to the surface of its event
horizon.}, the exact numerical factor has not yet been determined
experimentally and thus might be questioned. For a determination
of this factor it is good know what thermodynamic relations in the
interior holostar space-time are independent from the Hawking
formula. The following derivations only make use of equation
(\ref{eq:FS:FE}), i.e.  $F_E = F_S$.

Knowing $u_F$ from equation (\ref{eq:u2}) the entropy per particle
$\sigma$ can be easily calculated by equations (\ref{eq:SF},
\ref{eq:NF}):

\begin{equation} \label{eq:SperN}
\sigma = \frac{\delta S}{\delta N} = \frac{F_S}{F_N} =
\frac{F_E}{F_N}
\end{equation}

with

\begin{equation} \label{eq:FN}
F_N(u_F) = f_F (Z_{F,2}(u_F) + Z_{F,2}(-u_F))+ 2 f_B Z_{B,2}(0)
\end{equation}

The energy per relativistic particle is given by equations
(\ref{eq:E_th}, \ref{eq:N:FN}). We find, just as in the previous
section, that the mean particle energy per temperature is constant
and equal to the mean entropy per particle:

\begin{equation} \label{eq:EperN}
\epsilon = \frac{\delta E}{\delta N} = \frac{F_E}{F_N} T = \sigma
T
\end{equation}

\begin{figure}[ht]
\begin{center}
\includegraphics[width=12cm, bb=60 433 500 767]{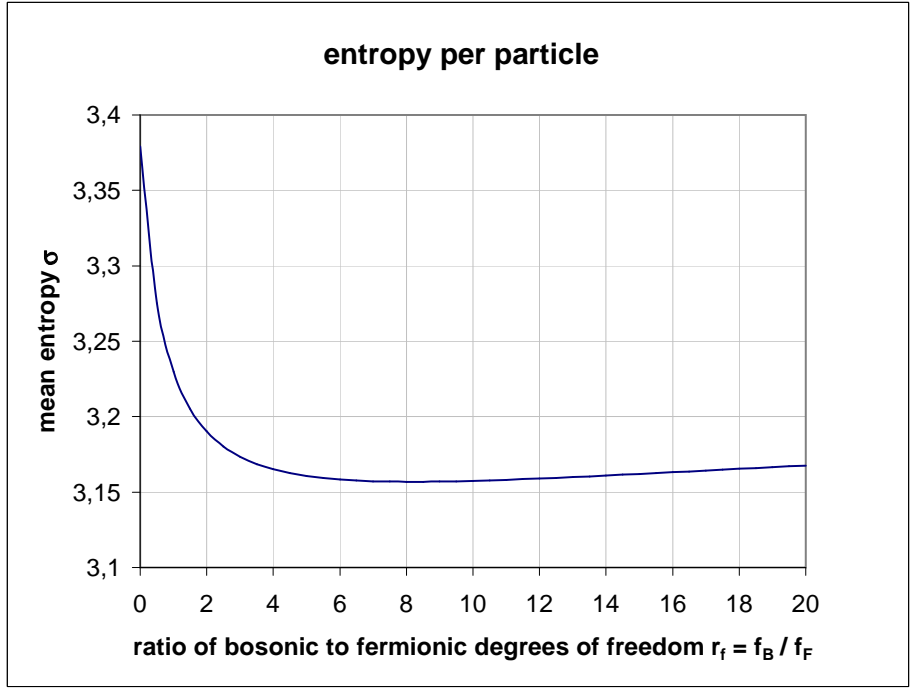}
\caption{\label{fig:epp}mean entropy per particle of the
ultra-relativistic fermions in the holographic solution as a
function of the ratio of bosonic to fermionic degrees of freedom
$r_f = f_B / f_F$}
\end{center}
\end{figure}

$\sigma$ only depends on the number of degrees of freedom of the
ultra-relativistic bosons and fermions in the model. In fact,
$\sigma$ only depends on the ratio $r_f = f_B/f_F$ and is a very
slowly varying function of this ratio. Figure \ref{fig:epp} shows
the dependence of $\sigma$ on $r_f$.

The relation $\epsilon = \sigma T$, which relates the mean energy
per particle to the mean entropy times the local radiation
temperature can be viewed as the {\em fundamental thermodynamic
characteristic} of the holostar. Keep in mind that this relation
is only valid for the {\em mean} energy per particle and the {\em
mean} entropy per particle, evaluated with respect to all
particles. It isn't fulfilled for the bosonic and fermionic
species individually. In general, except for the special case $f_B
= 0$, we have $\epsilon_B \neq \sigma_B T$ and $\epsilon_F \neq
\sigma_F T$.

The relation $\epsilon = \sigma T$, which is equivalent to $F_S =
F_E$, has the remarkable side-effect, that the free energy is
identical zero in the holostar solution:

\begin{equation} \label{eq:F}
F = E - S T = N (\epsilon - \sigma T ) = 0
\end{equation}

Usually a closed system has the tendency to minimize it's free
energy, which is a compromise between minimizing it's energy and
maximizing it's entropy. The holostar is the prototype of a closed
system. It is a self-gravitating static solution to the Einstein
field equations. It's only form of energy-exchange with the
exterior space-time is through Hawking-radiation, which is an
utterly negligible mode of energy-exchange for a large holostar.
In this respect it is remarkable that the holostar solution
minimizes the free energy to zero, e.g. the smallest possible
value that a sensible measure of energy in general relativity can
have. This indicates, that the free energy in general relativity
might be more than a mere book-keeping device.

With the help of equation (\ref{eq:FS:FE}), but not using equation
(\ref{eq:FE:beta}), the entropy within the shell can be expressed
as:

\begin{equation} \label{eq:delta:S}
\delta S(r) = \left(\frac{F_E}{4 \pi
\beta}\right)^{\frac{1}{4}}\frac{r \delta r }{\hbar}
\end{equation}

If the total entropy of the holostar, i.e. the integral over the
entropy-contributions of the respective shells, is to be
proportional to the Hawking entropy of a black hole with the same
gravitational radius, $F_E / \beta$ must be constant. Integration
of equation (\ref{eq:delta:S}) gives the result:

\begin{equation} \label{eq:S:unfixed}
S = \frac{1}{2\pi} \left(\frac{F_E}{4 \pi \beta}\right)^{\frac{1}{4}} \frac{A}{4 \hbar}
\end{equation}

The Hawking result is reproduced, whenever:

\begin{equation} \label{eq:kappa}
\omega = \frac{1}{2 \pi} \left(\frac{F_E}{4 \pi
\beta}\right)^{\frac{1}{4}} = 1
\end{equation}

$\omega$, which depends on the ratio $F_E / \beta$, is the
constant of proportionality between the holostar entropy and the
Hawking entropy. Setting $\omega = 1$ is equivalent to equation
(\ref{eq:FE:beta}), which fixes $\beta$ with respect to the
Hawking temperature. If the Hawking entropy/temperature formula
have to be rescaled, $\omega$ is nothing else than the (nearly
constant) scale factor. Therefore let us express all thermodynamic
relations in terms of $\omega$.

The number of particles within the shell is given by equation
(\ref{eq:NF}), which is extended to encompass the bosonic degrees
of freedom:

\begin{equation} \label{eq:N:FN}
\delta N(r) = \frac{F_N}{F_E} \left(\frac{F_E}{4 \pi
\beta}\right)^{\frac{1}{4}} \frac{r \delta r}{\hbar} =
\frac{\omega}{\sigma} \frac{2 \pi r \delta r}{\hbar}
\end{equation}

The total number of particles is given by a simple integration,
assuming that $\omega = const$:

\begin{equation} \label{eq:N:holostar}
N = \left(\frac{F_E}{4 \pi \beta}\right)^{\frac{1}{4}} \frac{1}{2
\pi} \frac{1}{\sigma} \frac{A}{4 \hbar} = \frac{\omega}{\sigma}
\frac{A}{4 \hbar}
\end{equation}

Therefore, as derived in the previous section, the total number of
particles within the holostar is proportional to its surface area,
whenever $F_E / \beta = const$ and $\sigma = const$.

The temperature of the holostar at infinity is given by

\begin{equation} \label{eq:T:inf:unfixed}
T_{\infty} = T(r_h) \sqrt{g_{tt}(r_h)} = 2\pi \left(\frac{4 \pi
\beta}{F_E }\right)^{\frac{1}{4}} \frac{\hbar}{4 \pi r_h} =
\frac{1}{\omega} \frac{\hbar}{4 \pi r_h}
\end{equation}

Again, if we set $\omega = 1$ we get the Hawking temperature. The
important result is, that $\omega$ could in principle take on any
arbitrary (nearly constant) value. This is possible, because the
factor in the temperature is just the inverse as the factor in the
entropy. As is well known from black hole physics, any constant
rescaling of the Hawking entropy must necessarily rescale the
temperature such, that the product of temperature and entropy is
equal for the scaled and unscaled quantities, i.e. $S \, T$ must
be unaffected by the rescaling. This is necessary, because
otherwise the thermodynamic identity

$$\frac{\partial S}{\partial E} T = 1$$

would not be fulfilled in the exterior space-time. (In the
exterior space-time the energy $E$ is fixed and is taken to be the
gravitating mass $M = r_h/2$ of the black hole.)

As can be seen from equations (\ref{eq:S:unfixed},
\ref{eq:T:inf:unfixed}), entropy and temperature at infinity of
the holostar fulfill the rescaling condition. Furthermore, entropy
and temperature at infinity are exactly proportional to the
Hawking temperature and entropy. This result is not trivial. It
depends on the holostar metric, which has just the right value at
the position of the membrane, so that the temperature at infinity
scales correctly with respect to the entropy.

\subsection{Relating the local thermodynamic temperature to the Hawking temperature}

Now we are ready to set $\omega=1$, which gives us the desired
relation between $\beta$ and $F_E$, as already expressed in
equation (\ref{eq:FE:beta}).

With $\omega = 1$, the local thermodynamic temperature of any
interior shell can be expressed solely in terms of $\beta$. It
turns out to be equal to the expression in equation
(\ref{eq:Tlocal}) of the previous section:

\begin{equation} \label{eq:T4local}
T^4 = \frac{\hbar^3}{(4 \pi)^4 \beta} \frac{1}{r^2}
\end{equation}

or

$$T^4 A = \frac{1}{\beta} \left(\frac{\hbar}{(4 \pi)} \right)^3 = const$$

\section{A measurement of the Hawking temperature}

In the previous section the internal temperature of the holostar
has been derived by "fixing" it with respect to the Hawking
temperature. Although Hawking's calculations are robust
\cite{Hawking/75} and there appears to be no reason, why the
Hawking equation should be modified - at least for large black
holes\footnote{The only ingredient in Hawking's derivation is the
propagation of a quantum field in the exterior vacuum space-time
of a black hole. Both concepts (quantum field in vacuum; exterior
space-time of a black hole) are very accurately understood.} - it
has been speculated whether the factor in the entropy-area law (or
in the temperature formula) might take a different value. A single
measurement of the Hawking temperature (or entropy) of a large
black hole could settle the question. However, with no black hole
available in our immediate vicinity and taking into account the
extremely low temperatures of even comparatively small black
holes, there appeared to be no feasible means to measure the
Hawking entropy or temperature of a black hole directly or
indirectly.

It would be of high theoretical value, if the Hawking
temperature/entropy formula could be verified (or falsified) by an
explicit measurement. The holostar provides such a means.

For this purpose let us assume, that the Hawking temperature formula
were modified by a constant factor, i.e

\begin{equation}
T = \frac{1}{\omega} \frac{\hbar}{4 \pi r}
\end{equation}

where $\omega$ is a dimensionless factor, whose value can be
determined experimentally.

If we set the temperature of the holostar equal to the modified
Hawking temperature we get the following result for $F_E$:

\begin{equation}
\frac{F_E}{4 \pi \beta} = \left(2 \pi \omega \right)^4
\end{equation}

The local temperature within the holostar is then given by
equation (\ref{eq:TlocTherm}):

\begin{equation}
T^4 = \frac{\hbar^3}{2^8 \pi^4 \beta r^2 \omega^4} =
\frac{1}{\omega^4} \frac{\hbar^3}{2^5 \pi^3 \beta} \rho
\end{equation}

$\rho = 1/(8\pi r^2)$ is the total (local) energy density of the
matter within the holostar. The above equation can be solved for
$\omega$:

\begin{equation}
\omega^4 = \frac{\hbar^3} {2^5 \beta \pi^3} \frac {\rho} {T^4}
\end{equation}

The local radiation temperature $T$ and the total local energy
density $\rho$ within a holostar are both accessible to
measurement. Note that the local temperature within a holostar is
much easier to measure than its (Hawking) temperature at infinity:
The local interior temperature only scales with $1/\sqrt{M}$,
whereas the temperature at infinity scales with $1/M$. Therefore
even a very large holostar will have an appreciable interior local
radiation temperature, although its Hawking temperature at
infinity will be unmeasurable by all practical means.

In order to determine $\omega$ the value of $\beta$ need to be
known. In \cite{petri/charge} the following formula for $\beta$
has been suggested:

\begin{equation} \label{eq:beta:running}
\frac{\beta}{4} = \frac{\alpha}{2} + \sqrt{\left(
\frac{\alpha}{2}\right)^2 + \frac{3}{4}}
\end{equation}

$\alpha$ is the running value of the fine-structure constant,
which depends on the local energy scale. Note that the above
relation for $\beta$ has not been derived rigorously in
\cite{petri/charge}, but was suggested by analogy, i.e. by
extrapolating the (exact) relation between mass, charge, boundary
area and $r_0$ derived for an extremely charged holostar to the
charged/rotating case. Angular momentum was introduced in
straightforward way, giving the correct formula for a Kerr-Newman
black hole in the macroscopic limit and the correct formula for
the a charged, non-rotating holostar for $J=0$. The formula with
non-zero $J$ then was applied to a {\em microscopic} object, a
spin-1/2 extremely charged holostar of minimal mass, in order to
obtain equation (\ref{eq:beta:running}). One must keep in mind
though, that in principle there are several ways to extend the
formula to the rotating case giving the correct macroscopic limit,
but which might differ in their microscopic predictions.

We wan't to apply equation (\ref{eq:beta:running}) in order to
derive the value of $r_0$ which determines the {\em interior}
radial metric coefficient of a large holostar, $g_{rr} = r_0/r$.
The implicit assumption which lies at the heart of equation
(\ref{eq:beta:running}), is that $r_0^2 = \beta \hbar$ is a {\em
universal} quantity, not dependent on the nature of the system in
question and only - moderately - dependent on the energy-scale. It
requires quite a leap of faith to do this. It is not possible to
fully justify this assumption in the context of this paper. See
\cite{petri/thermo, petri/charge} for a more detailed discussion.

Due to the appearance of $\alpha$ in the formula for $r_0^2$ it is
suggestive to interpret $r_0$ as a running length scale, which
depends on the energy $E$ via $\alpha(E)$. This means that for
high temperatures $r_0$ is expected to increase with energy as a
function of $\alpha(E)$. This makes sense, because we have already
seen, that $r_0^2$ is proportional to the effective degrees of
freedom, which are also known to increase at high energies.
Therefore, if we treat $r_0^2$ as a universal quantity, the only
sensible way is to interpret $\alpha$ as the running value of the
relevant coupling constants depending on the energy scale.

With this interpretation, whenever $\alpha$ is small, such as for
the typical energies encountered today, it can be set to zero in
the above equation to a very good approximation, so that $\beta
\approx 4 \sqrt{3/4}$.

Let us now make the assumption, that we live in a large holostar.
In \cite{petri/hol} several observational facts have been
accumulated which suggest that such a claim is not too far
fetched. Then the local radiation temperature will be nothing else
than the microwave-background temperature and the total (local)
energy density will be the total matter density of the universe at
the present time (= present radial position). Both quantities have
been determined quite precisely in the recent past. With the
following value for the temperature of the microwave background
radiation

$$T_{CMBR} = 2.725 K$$

and with the total matter density determined from the recent
WMAP-measurements \cite{WMAP/cosmologicalParameters}

$$\rho = 0.26 \, \rho_{c} = 2.465 \cdot 10^{-27}
\frac{kg}{m^3}$$

and with $\beta$ determined from equation
({\ref{eq:beta:running}}) using the present (low energy) value of
the fine-structure constant, $\alpha$,

$$\beta = 3.479$$

we find:

\begin{equation}
\omega^4 = 1.0116
\end{equation}

or

$$\omega = 1.003$$

If we set the fine-structure constant to zero, i.e. $\beta = 4
\sqrt{3/4}$, the agreement is almost as good: $\omega = 1.004$.
The very high accuracy suggested in the above results is somewhat
deceptive. With $T$ known to roughly 0.1\% the error in $\omega$
will be dominated by the uncertainty in $\rho$. A conservative
estimate for this uncertainty should be roughly 5\%. Taking the
fourth square root suppresses the relative error by roughly a
factor of four, so that the error in $\omega$ will be roughly 1\%.
Therefore, within the uncertainties of the determination of $\rho$
and $T$ the Hawking-entropy formula is reproduced to a remarkably
high degree of accuracy of roughly 1\%.

Not knowing $\beta$, the experimental data only allow us to
determine $\omega^4 \beta \approx 3.519$. Therefore, as long as
equation (\ref{eq:beta:running}) has not been verified
independently, it is prudent to keep this caveat in mind.

\section{Matter-dominated holostars}

So far we have assumed, that the interior matter-state consists of
an ultra-relativistic gas. At low temperatures, well below the
rest-masses of the fundamental particles this is not the case. Is
a matter-dominated holostar also compatible with the Hawking
entropy and temperature?

The entropy of a massive particle (with zero chemical potential)
is given by:

\begin{equation} \label{eq:s:m}
\sigma_m = \frac{m}{T}
\end{equation}

As long as there is at least one relativistic particle left, we
will have a radiation temperature given by equation
(\ref{eq:Tlocal}). Let us assume, that the matter is dominated by
one massive particle species, such as the nucleon. Then the
number-density of the massive particles is simply given by:

\begin{equation} \label{eq:n:m}
n_m = \frac{\rho}{m} = \frac{1}{8 \pi r^2 \, m}
\end{equation}

The local entropy-density is given by the product of equations
(\ref{eq:s:m}, \ref{eq:n:m}) and is independent of particle mass.

\begin{equation}
s = n_m \sigma_m = \frac{1}{8 \pi r^2} \frac{1}{T} = \frac{1}{2 r
\hbar} \sqrt{\frac{r_0}{r}}
\end{equation}

Therefore the above result equally applies to a mixture of
particles with different masses. The total entropy follows from a
proper integration over the interior entropy-density:

\begin{equation} \label{eq:S:m}
S = \int_0^{r_h}{s dV} = \frac{\pi r_h^2}{\hbar} = S_{BH}
\end{equation}

\section{Discussion and Outlook}

A simple thermodynamic model for a compact self-gravitating object
with an interior matter-density $\rho = 1 / (8 \pi r^2)$ has been
presented which fits well into the established theory of black
holes. From the viewpoint of an exterior observer the object
appears very similar to a classical black hole. The modifications
are minor and only "visible" at close distance:

The event horizon is replaced by a two dimensional membrane with
high tangential pressure, situated roughly two Planck coordinate
lengths outside of the object's gravitational radius. The surface
redshift at the membrane scales with $z = \sqrt{r/r_0}$, where
$r_0$ is a fundamental length, roughly equal to two Planck
lengths. A solar mass object has a surface redshift $z\approx
10^{20}$.

Simply by assuming (i) spherical symmetry, (ii) Einstein's field
equations with zero cosmological constant and (iii) microscopic
statistical thermodynamics in the ideal gas approximation it could
be shown that any compact self-gravitating object with an interior
matter-density $\rho = 1 / (8 \pi r^2)$ has a thermodynamic
entropy and a temperature at infinity exactly proportional to the
Hawking entropy and -temperature. The number of interior
ultra-relativistic particles is proportional to the proper area of
the object's boundary-membrane, measured in Planck units,
indicating that the holographic principle is valid for compact
self-gravitating objects of arbitrary size.

The object has a well-defined interior temperature with $T \propto
1 / \sqrt{r}$. The object's surface temperature can be related to
the Hawking temperature. By this correspondence one can set up a
specific relation between the Hawking temperature (measured at
infinity), the interior radiation temperature and the interior
matter density. This correspondence allows an experimental
verification of the Hawking-temperature law, by measurements in
the object's interior.

At ultra-relativistic energies the fermions acquire a non-zero
chemical potential. The chemical potential per temperature $u =
\mu / T$ can be calculated by a closed formula. Its value depends
only on the ratio of bosonic to fermionic degrees of freedom and
is a monotonically increasing function of this ratio. If there are
only fermions, $u \approx 1.34$. In the supersymmetric case (equal
fermionic and bosonic degrees of freedom) $u = \pi / \sqrt{3}
\approx 1.8$. The non-zero chemical potential of the fermions
naturally induces a profound matter-antimatter asymmetry at high
temperatures. The implications of this finding are discussed in
\cite{petri/asym, petri/thermo}.

One particularly interesting solution to the field equations with
an interior matter-density $1/ (8\pi r^2)$ is the so called
holographic solution, short holostar. Its remarkable properties
have been discussed extensively in \cite{petri/hol, petri/charge,
petri/string, petri/thermo}. The holostar's membrane has a
pressure equal to the pressure derived from the so called
"membrane paradigm" for black holes \cite{Thorne/mem,
Price/Thorne/mem}. This guarantees, that the holostar's action on
the exterior space-time is practically indistinguishable from that
of a same-sized black hole. The membrane has zero energy-density,
as expected from string theory. Its interior matter has an overall
string equation of state. The strings are densely packed, each
string occupying a transverse extension of exactly one Planck
area. This dense package of strings is the fundamental reason why
the holographic solution does not collapse to a singularity,
although its membrane lies barely two Planck coordinate distances
outside its gravitational radius.

The holostar solution has no singularity and no event horizon.
Information is not lost: The total information content of the
space-time is encoded in its constituent matter, which can consist
out of strings or particles. Unitary evolution of particles is
possible throughout the full space-time manifold. Every
ultra-relativistic particle carries a definite entropy, which can
be calculated when the ratio of bosonic to fermionic degrees of
freedom is known.

The holostar solution has been shown in \cite{petri/hol} to be an
astoundingly accurate model for the universe, as we see it today.
By comparing the CMBR-temperature to the total matter density of
the universe as determined by WMAP the Hawking temperature law has
been experimentally verified to an accuracy of roughly 1 \%.
However, the exact numerical verification depends on an equation
which has been suggested by analogy in \cite{petri/charge}, but
stills lacks a formal derivation.

Having two or more solutions for the field equations (black hole
vs. holostar) makes the question of how these solutions can be
distinguished from each other experimentally an imminently
important question. Can we find out by experiment or observation,
which of the known solutions, if any, is realized in nature? At
the present time the best argument in favor of the holostar
solution appears to be the accurate measurement of the Hawking
temperature via the CMBR-temperature and the matter-density of the
universe.

Yet it would be helpful if more direct experimental evidence were
available. Due to Birkhoff's theorem the holostar cannot be
distinguished from a Schwarzschild black hole by measurements of
its exterior gravitational field. But whenever holostars come
close to each other or collide, their characteristic interior
structure should produce observable effects, which deviate from
the collisions of black holes. Presumably a collision of two
holostars will be accompanied by an intense exchange of particles,
with the possible production of particle jets along the angular
momentum axis.

In accretion processes the membrane might produce a noticeable
effect. The rather stiff membrane with its high surface pressure
might be a better "reflector" for the incoming particles, than the
vacuum-region of the event horizon of a Schwarzschild-type black
hole. There are observations of burst-like emissions from compact
objects, which are assumed to be black holes because of their high
mass ($M > 3-5 M_{\odot}$), but that have "hard" spectra rather
characteristic for neutron stars. A more accurate observation of
these objects might provide important experimental clues to decide
the issue.

For holostars of sub-stellar size ($r_h \approx 1 \, km)$ the
local temperature at the membrane becomes comparable to the
nucleon rest mass energy. A rather hot particle gas at the
position of the membrane could produce noticeable effects with
respect to the relative abundances of the "reflected" particles,
due to high energy interactions with the membrane or the
holostar's interior.

On the other hand, the extreme surface red-shifts on the order of
$z \approx 10^{20}$ for a solar mass holostar, and larger yet for
higher mass objects ($z \propto \sqrt{M}$), might not allow a
conclusive interpretation of the experimental data with regard to
the true nature of any such black hole type object.

The most promising route therefore appears to be, to study the
holostar from its interior. In \cite{petri/hol} it has been
demonstrated, that the holostar has the potential to serve as an
alternative model for the universe. The recent WMAP-measurements
have determined the product of the Hubble constant $H$ times the
age of the universe $\tau$ to be $H \, \tau \simeq 1.02$
experimentally with $H = 71 \, (km/s)/Mpc$ and $\tau = 13.7 \,
Gy$. The holostar solution predicts $H \, \tau = 1$ exactly. There
are other predictions which fit astoundingly well with the
observational data. This in itself is remarkable, because the
holostar-solution has practically no free parameters. It's unique
properties arise from a delicate cancelation of terms in the
Einstein field equations, which only occurs for the "special"
matter density $\rho = 1 / (8 \pi r^2)$ in combination with a
string equation of state, leading to the "special" radial metric
coefficient $g_{rr} = r/r_0$ and a time coefficient $g_{tt} =
r_0/r$. That the holostar solution with its completely "rigid"
structure has so much in common with the universe as we see it
today, either is the greatest coincidence imaginable, or not a
coincidence at all.

With the holostar solution we have a beautifully simple model for
a singularity free compact self gravitating object, which is
easily falsifiable. Its metric and fields are simple, its
properties are not. It is an elegant solution, as anyone studying
its properties will soon come to realize. However, in science it
is experiments and observations, not aesthetics, that will have to
decide, which solution of the field equations has been chosen by
nature. It is our task, to find out. The work has just begun.


\end{document}